%Paper: 9204013
%From: hodgdon@helios.TN.CORNELL.EDU (Jennifer Hodgdon)
%Date: Fri, 24 Apr 92 10:50:24 EDT

%%%%%%%%%%%%%%%%%%%%main.tex%%%%%%%%%%%%%%%%%%%%%%%%%%
%  NOTE: there are several files appended to this main tex file
%        which you will need to separate from the main file before
%        tex-ing the file.  The first is a non-standard macro package,
%        which I have called jensmac.tex; the rest are postscript files
%	 for the figures.  The files are all separated by rows
%        of percent signs, with the file names imbedded.  The standard
%        macro packages psfig.tex and tables.tex are also used by this
%        tex file.  If you do not want the figures, you will need to
%        discard the .ps files at the end and remove all the psfig
%        commands in this main.tex file.
%
%%%%%%%%%%%%%%%%%%%%jensmac.tex%%%%%%%%%%%%%%%%%%%%%%%%%%%%%%%%%%%%%%%%

% I have removed most of the comments from these macros, for space.

\def\today{\ifcase\month\or
  January\or February\or March\or April\or May\or June\or
  July\or August\or September\or October\or November\or December\fi
  \space\number\day, \number\year}
\def\prtoday{\number\day\space\ifcase\month\or
  January\or February\or March\or April\or May\or June\or
  July\or August\or September\or October\or November\or December\fi
  \space\number\year}
\def\draftid{\relax}
\def\reftag{\relax}
\def\equtag{\relax}
\def\figtag{\relax}
\def\tabtag{\relax}
\def\hideunder#1#2{% hide #1 directly underneath #2
    \oalign{{#2}\crcr\hidewidth{#1}\hidewidth}}
\normalbaselineskip= 12.4pt
\def\singlespace{\baselineskip=\normalbaselineskip}
\def\oneandahalfspace{\baselineskip=\normalbaselineskip
                      \multiply \baselineskip by 3
                      \divide \baselineskip by 2}
\def\doublespace{\baselineskip=\normalbaselineskip
                 \multiply \baselineskip by 2}
\def\skipline{\vskip\baselineskip}
        \magnification=\magstep1
	\parskip=0pt plus6pt
	\doublespace
   	\hsize   = 5.750truein
	\vsize   = 8.750truein
	\hoffset = 0.500truein
	\voffset = 0.125truein
\overfullrule = 0pt
\raggedbottom
\newif\ifnewfoot
\newif\ifnewhead
\def\newchapter#1{%
	\def\secname{#1}
	\global\newfoottrue
	\global\newheadtrue
	\def\refkeys{}
	\def\equkeys{}
	\def\figkeys{}
	\def\tabkeys{}
	\def\equlist{}
	\def\figlist{}
	\def\tablist{}
	\headline={\ifnewhead\newheader\global\newheadfalse \else\oldheader\fi}
	\footline={\ifnewfoot\newfooter\global\newfootfalse \else\oldfooter\fi}
        }
\def\newheader{\hfil}
\def\newfooter{\centerpagenum}
\def\oldheader{\centerpagenum}
\def\oldfooter{\hfil}
\def\centerpagenum{\hss\tenrm\folio\hss\draftid}
\newchapter{}
\def\equkeys{}%		the list of keywords; accumulated as cited
\def\equlist{}%		the list of keywords twice.
\def\equn#1{\buildlist{#1}\updateequs{\listed}\citeequs{\listed}}
\def\updateequs#1{\citenum=0%
                 \def\\##1{\updateequ{##1}}%
                 #1}%
\def\updateequ#1{\advance\citenum by 1%
                 \ifx\equkeys\empty%	For the first citation, start the list
                    \add{#1}\to{\equkeys}%
                    \add{#1}\to{\equlist}%
                    \add{#1}\to{\equlist}%
                 \else%
                    \memberfalse%
                    \ismember{#1}\of{\equkeys}%	check if key is on list
                    \def\\##1{\updateequ{##1}}% reset the \\ definition, which
                    \ifmember%			gets changed during \ismember
                    \else%
                       \add{#1}\to{\equkeys}%	If key isn't on list, add it
                       \add{#1}\to{\equlist}%
                       \add{#1}\to{\equlist}%
                    \fi%
                 \fi}%
\def\citeequs#1{\refnum=0%
                \seqnum=0%
                \hideunder{\equtag}{\secname
                \def\\##1{\citedef{#1}{##1}}%
                \equkeys%
                \def\\##1{}}}
\long\def\listequs{{\bf EQUATION NAMES:}
                   \refnum=0%
                   \def\\##1{\advance\refnum by 1%
                             \writeequ{\number\refnum}{##1}}%
                   \equkeys}
\long\def\writeequ#1#2{\memberfalse%
                       \def\nextkey{#2}%
                       \long\def\\##1\\##2{\def\nextbibkey{##1}%
                                           \ifmember
                                           \else
                                              \ifx\nextkey\nextbibkey
                                                 \membertrue
                                                 \printequ{#1}{##2}
                                              \fi
                                           \fi}%
                       \equlist%
                       \def\\##1{\advance\refnum by 1%
                                 \writeequ{\number\refnum}{##1}}}
\long\def\printequ#1#2{%
               \begingroup
               \par
               \noindent
               \advance\leftskip by 1truein
               \llap{Eq.\ (\secname#1)\ =\ }#2
               \endgroup}

\def\figkeys{}%		the list of keywords; accumulated as cited
\def\figlist{}%		the list of keywords and text; built by \figitem

\def\figitem#1#2{\add{#1}\to{\figlist}%
                 \add{#2}\to{\figlist}}
\def\updatefigs#1{\citenum=0%
                 \def\\##1{\updatefig{##1}}%
                 #1}%
\def\updatefig#1{\advance\citenum by 1%
                 \ifx\figkeys\empty%	For the first citation, start the list
                    \add{#1}\to{\figkeys}%
                 \else%
                    \memberfalse%
                    \ismember{#1}\of{\figkeys}%	check if key is on list
                    \def\\##1{\updatefig{##1}}% reset the \\ definition, which
                    \ifmember%			gets changed during \ismember
                    \else%
                       \add{#1}\to{\figkeys}%	If key isn't on list, add it
                    \fi%
                 \fi}%
\def\citefigs#1{\refnum=0%
                \seqnum=0%
                \hideunder{\figtag}{\secname
                \def\\##1{\citedef{#1}{##1}}%
                \figkeys%
                \def\\##1{}}}
\long\def\listfigs{\refnum=0%
                   \def\\##1{\advance\refnum by 1%
                             \writefig{\number\refnum}{##1}}%
                   \figkeys}
\long\def\writefig#1#2{\memberfalse%
                       \def\nextkey{#2}%
                       \long\def\\##1\\##2{\def\nextbibkey{##1}%
                                           \ifmember
                                           \else
                                              \ifx\nextkey\nextbibkey
                                                 \membertrue
                                                 \printfig{#1}{##2}
                                              \fi
                                           \fi}%
                       \figlist%
                       \def\\##1{\advance\refnum by 1%
                                 \writefig{\number\refnum}{##1}}}
\long\def\printfig#1#2{%
               \begingroup
               \bigskip
               \noindent
               \singlespace
               \advance\leftskip by 1truein
               \llap{\bf Fig.\ #1\ \ }#2\par
               \endgroup}

\def\figcap#1{\buildlist{#1}\updatefigs{\listed}\listonefig}
\long\def\listonefig{\def\\##1{%
                           \writefig{\citefigs{\listed}}{##1}}%
                     \listed}
\def\tabkeys{}%		the list of keywords; accumulated as cited
\def\tablist{}%		the list of keywords and text; built by \tabitem

\def\tabitem#1#2{\add{#1}\to{\tablist}%
                 \add{#2}\to{\tablist}}
\def\updatetabs#1{\citenum=0%
                 \def\\##1{\updatetab{##1}}%
                 #1}%
\def\updatetab#1{\advance\citenum by 1%
                 \ifx\tabkeys\empty%	For the first citation, start the list
                    \add{#1}\to{\tabkeys}%
                 \else%
                    \memberfalse%
                    \ismember{#1}\of{\tabkeys}%	check if key is on list
                    \def\\##1{\updatetab{##1}}% reset the \\ definition, which
                    \ifmember%			gets changed during \ismember
                    \else%
                       \add{#1}\to{\tabkeys}%	If key isn't on list, add it
                    \fi%
                 \fi}%
\def\citetabs#1{\refnum=0%
                \seqnum=0%
                \hideunder{\tabtag}{\secname
                \def\\##1{\citedef{#1}{##1}}%
                \tabkeys%
                \def\\##1{}}}
\long\def\listtabs{\refnum=0%
                   \def\\##1{\advance\refnum by 1%
                             \writetab{\number\refnum}{##1}}%
                   \tabkeys}
\long\def\writetab#1#2{\memberfalse%
                       \def\nextkey{#2}%
                       \long\def\\##1\\##2{\def\nextbibkey{##1}%
                                           \ifmember
                                           \else
                                              \ifx\nextkey\nextbibkey
                                                 \membertrue
                                                 \printtab{#1}{##2}
                                              \fi
                                           \fi}%
                       \tablist%
                       \def\\##1{\advance\refnum by 1%
                                 \writetab{\number\refnum}{##1}}}
\long\def\printtab#1#2{%
               \begingroup
               \bigskip
               \noindent
               \singlespace
               \advance\leftskip by 1truein
               \llap{\bf Table\ #1\ \ }#2\par
               \endgroup}
\def\tabinsert#1#2{\topinsert\oneandahalfspace
                   #1
                   \medskip
                   \tabcap{#2}
                   \medskip
                   \endinsert}
\def\tabcap#1{\buildlist{#1}\updatetabs{\listed}\listonetab}
\long\def\listonetab{\def\\##1{%
                           \writetab{\citetabs{\listed}}{##1}}%
                     \listed}

\def\IJF{Int.\ J.\ Fracture }

\def\JMS{J.\ Mat.\ Sci.\ }

\def\PRA{Phys.\ Rev.\ A}

\def\PRL{Phys.\ Rev.\ Lett.\ }
\def\RMP{Rev.\ Mod.\ Phys.\ }

\def\refkeys{}%		the list of keywords; accumulated as cited
\def\biblist{}%		the list of keywords and text; built by \bibitem
\def\aylist{}%          the list of author/title citations
\newtoks\ta
\newtoks\tb
\newif\ifmember
\newif\ifauthoryear
\authoryearfalse
\newif\iftitlesin
\titlesinfalse
\newcount\citenum%	the number of items in a given \cite command
\newcount\refnum%       the reference number
\newcount\seqnum%       the number in sequence
\newcount\seqfirst%     the first in a sequence
\newcount\seqlast%      the last in a sequence

\def\cite#1{\buildlist{#1}\updaterefs{\listed}$^{\citerefs{\listed}}$}
\def\pcite#1{\cite{#1}\hskip 4.4443pt plus 4.99997pt minus 0.37036pt}
\def\refnumber#1{\buildlist{#1}\updaterefs{\listed}\citerefs{\listed}}
\def\quietcite#1{\buildlist{#1}\updaterefs{\listed}}
\def\jref#1#2#3#4#5#6#7#8{%
  \iftitlesin\bibitem{#1}{#2, ``#8,'' #3 {\bf #4}, #5 (#6).}%
  \else\bibitem{#1}{#2, #3 {\bf #4}, #5 (#6).}%
  \fi%
  \ifauthoryear\quietcite{#1}%
                \add{#1}\to\aylist%
                \add{#7, #6}\to{\aylist}%
  \fi }
\def\bref#1#2#3#4#5{%
  \bibitem{#1}{#2, #3  (#4).}%
  \ifauthoryear\quietcite{#1}%
                \add{#1}\to\aylist%
                \add{#5, #4}\to{\aylist}%
  \fi }
\def\pref#1#2#3#4#5#6{%
  \bibitem{#1}{#2, ``#6,'' #3  (#4).}%
  \ifauthoryear\quietcite{#1}%
                \add{#1}\to\aylist%
                \add{#5, #4}\to{\aylist}%
  \fi }
\def\aref#1#2#3#4#5#6#7{%
  \iftitlesin\bibitem{#1}{#2, ``#7,'' p. #3 in #4 (#5).}%
  \else\bibitem{#1}{#2, p. #3 in #4 (#5).}%
  \fi%
  \ifauthoryear\quietcite{#1}%
                \add{#1}\to{\aylist}%
                \add{#6, #5}\to{\aylist}%
  \fi }
\def\bibitem#1#2{\add{#1}\to{\biblist}%
                 \add{#2}\to{\biblist}}
\long\def\add#1\to#2{\ta={\\{#1}}%
                     \tb=\expandafter{#2}%
                     \global\edef#2{\the\tb\the\ta}}%
\def\buildlist#1{\def\listed{}\makelistmacro#1,@}
\def\makelistmacro#1#2,#3@{\add{#1#2}\to{\listed}%
                         \def\restoflist{#3}%
                         \ifx\restoflist\empty\relax%
                         \else\makelistmacro#3@%
                         \fi}
\def\updaterefs#1{\citenum=0%
                 \def\\##1{\updatedef{##1}}%
                 #1}%
\def\updatedef#1{\advance\citenum by 1%
                 \ifx\refkeys\empty%	For the first citation, start the list
                    \add{#1}\to{\refkeys}%
                 \else%
                    \memberfalse%
                    \ismember{#1}\of{\refkeys}%	check if key is on list
                    \def\\##1{\updatedef{##1}}% reset the \\ definition, which
                    \ifmember%			gets changed during \ismember
                    \else%
                       \add{#1}\to{\refkeys}%	If key isn't on list, add it
                    \fi%
                 \fi}%
\def\ismember#1\of#2{\def\given{#1}%
                     \def\\##1{\def\next{##1}%
                               \ifmember%
                               \else%
                                  \ifx\next\given%
                                      \membertrue%
                                  \fi%
                               \fi}%
                     #2}%
\def\citerefs#1{\refnum=0%
                \seqnum=0%
                \hideunder{\reftag}{\relax
                \def\\##1{\citedef{#1}{##1}}%
                \refkeys%
                \def\\##1{}}}
\def\citedef#1#2{\ifnum\citenum>0
                     \advance\refnum by 1%
                     \memberfalse%
                     \ismember{#2}\of{#1}%	was refkey cited?
                     \def\\##1{\citedef{#1}{##1}}%	redefine \\
                     \ifmember
                        \advance\citenum by-1 %
                        \ifnum\seqnum=0 %
                           \ifnum\citenum=0 %
                              \number\refnum %
                           \else %
                              \advance\seqnum by 1 %
                              \seqfirst=\refnum %
                           \fi %
                        \else %
                           \advance\seqnum by 1 %
                           \ifnum\citenum=0 %
                              \ifnum\seqnum=2 %
                                 \number\seqfirst ,\number\refnum %
                              \else %
                                 \number\seqfirst \hbox{--}\number\refnum %
                              \fi %
                           \else %
                              \seqlast=\refnum %
                           \fi %
                        \fi %
                     \else %
                        \ifnum\seqnum>2 %
                           \number\seqfirst \hbox{--}\number\seqlast ,%
                        \else %
                           \ifnum\seqnum=2 %
                              \number\seqfirst ,\number\seqlast ,%
                           \fi %
                           \ifnum\seqnum=1 %
                              \number\seqfirst ,%
                           \fi%
                        \fi%
                        \seqnum=0 %
                     \fi %
                  \fi}

\def\printiflisted#1#2{%
         \edef\given{#1}%
         \def\\##1{\edef\next{##1}%
                   \ifx\next\given%
                          #2%
                          \global\advance\citenum by -1%
                          \ifnum\citenum>1 ; \fi%
                   \fi}%
         \listed}

\long\def\listrefs{\refnum=0%
                   \def\\##1{\advance\refnum by 1%
                             \writeref{\number\refnum}{##1}}%
                   \refkeys}
\long\def\writeref#1#2{\memberfalse%
                       \def\nextkey{#2}%
                       \long\def\\##1\\##2{\def\nextbibkey{##1}%
                                           \ifmember
                                           \else
                                              \ifx\nextkey\nextbibkey
                                                 \membertrue
                                                 \printref{#1}{##2}
                                              \fi
                                           \fi}%
                       \biblist%
                       \def\\##1{\advance\refnum by 1%
                                 \writeref{\number\refnum}{##1}}}
\long\def\printref#1#2{%
               \begingroup
               \ifauthoryear
                   {#2}\par
               \else
                  \singlespace\noindent
                  \advance\leftskip by 0.5truein
                  \llap{#1.\ \ }{#2}\par
               \fi
          %     \skipline
               \endgroup}

\def\frac#1/#2{\leavevmode\kern.1em
  \raise.5ex\hbox{\the\scriptfont0 #1}\kern-.1em
  /\kern-.15em\lower.25ex\hbox{\the\scriptfont0 #2}}

\def\boxit#1{\vbox{\hrule\hbox{\vrule\kern3pt
  \vbox{\kern3pt#1\kern3pt}\kern3pt\vrule}\hrule}}

\def\hideunder#1#2{% hide #1 directly underneath #2
              \oalign{{#2}\crcr\hidewidth{#1}\hidewidth}}

\newif\ifnextcharnoskip
\def\noskipcharacters{\\{.}\\{,}\\{)}\\{:}\\{;}}
\def\nextcharcheck#1{\def\\##1{\checknoskip{#1}{##1}}%
                     \nextcharnoskipfalse%
                     \noskipcharacters%
                     \ifnextcharnoskip\else\ \fi%
                     \def\\##1{}%
                     #1}%
\def\checknoskip#1#2{\def\checkaa{#1}\def\checkbb{#2}%
                       \ifx\checkaa\checkbb%
                          \nextcharnoskiptrue%
                          \def\\##1{}%
                       \fi}%
\def\degC#1 {\mmmode{#1\ ^\circ{\rm C}}\nextcharcheck}
\def\K#1 {\mmmode{#1\ {\rm K}}\nextcharcheck}
\def\mK#1 {\mmmode{#1\ {\rm mK}}\nextcharcheck}
\def\Angstrom#1 {\mmmode{#1{\rm\ \AA}}\nextcharcheck}
\def\invAngstrom#1 {\mmmode{#1{\rm\ \AA}^{-1}}\nextcharcheck}
\def\nanometer#1 {\mmmode{#1\hbox{\ nm}}\nextcharcheck}
\def\micron#1 {\mmmode{#1\ \mu{\rm m}}\nextcharcheck}
\def\millimeter#1 {\mmmode{#1\hbox{\ mm}}\nextcharcheck}
\def\centimeter#1 {\mmmode{#1\hbox{\ cm}}\nextcharcheck}
\def\deg#1 {\mmmode{#1^\circ}\nextcharcheck}
\def\mrad#1 {\mmmode{#1\hbox{ mrad}}\nextcharcheck}
\def\rad#1 {\mmmode{#1{\rm\ rad}}\nextcharcheck}

\def\hours#1 {\mmmode{#1{\ \rm hrs}}\nextcharcheck}
\def\seconds#1 {\mmmode{#1{\ \rm sec.}}\nextcharcheck}
\def\Hz#1 {\mmmode{#1{\rm\ Hz}}\nextcharcheck}
\def\volts#1 {\mmmode{#1{\rm\ V}}\nextcharcheck}
\def\keV#1 {\mmmode{#1\hbox{\ keV}}\nextcharcheck}
\def\eV#1 {\mmmode{#1{\rm\ eV}}\nextcharcheck}
\def\mmmode#1{\ifmmode #1%
              \else $#1$%
              \fi}

% for holding up one end of a piece of glue
\def\scinot#1E#2 {\mmmode{#1\times 10^{#2}}}
\long\def\vmid#1{
  \bgroup
    \setbox100=\vbox{#1}
    \dimen100=\ht100
    \advance\dimen100 by \dp100
    \divide\dimen100 by 2
    \dimen101=\dimen100
    \advance\dimen100 by -\dp100
    \setbox101=\hbox{\lower\dimen100\copy100}
    \ht101=\dimen100
    \dp101=\dimen100
    \copy101
  \egroup}
\long\def\vupp#1{% returns #1 vertical list in a box
                 % of 0 ht and dp, and with reference point
                 % at top
  \bgroup
    \setbox100=\vbox{#1}
    \setbox101=\hbox{\lower\ht100\copy100}
    \ht101=0pt
    \dp101=0pt
    \copy101
  \egroup}
\long\def\vlow#1{% returns #1 in box with base at bottom
  \bgroup
    \setbox100=\vbox{#1}
    \setbox101=\hbox{\raise\dp100\copy100}
    \ht101=0pt
    \dp101=0pt
    \copy101
  \egroup}
 \def\topicname#1{\skipline\line{\bf#1\hfil}\nobreak}
 
\let\del=\nabla
\def\ddx#1#2{{\partial #2\over\partial #1}}
\def\dop#1{{\partial\over\partial #1}}

\def\putn#1{\eqno(\equn{#1})}

\def\half{{1 \over 2}}

\def\kI{K_{\rm I}}
\def\kII{K_{\rm II}}
\def\kIII{K_{\rm III}}
\def\nhat{\hat n}
\def\bhat{\hat b}
\def\that{\hat t}
\def\xvec{\vec x}

\def\dadt{{\partial a \over \partial t}}

\def\dds{{\partial \over \partial s}}

\def\dadotb#1#2{ {\partial #1 \over \partial s} \cdot #2 }

\input psfig
\input tables
\newchapter{}
%  BIBITEMS
\jref{bdl}{M.~Barber, J.~Donley, and J.~S. Langer}{\PRA}{40}{366}{1989}
{Barber}{Steady-state propagation of a crack in a viscoelastic strip}

\jref{BY}{K.~Binder and A.~P.~Young}{\RMP}{58}{801 (see pages 846-7)}
{1986}{Binder}{ TITLE????}

\bref{eefm}{David Broek}{{\bf Elementary Engineering Fracture Mechanics},
fourth edition; Nijhoff Publishers, Dordrecht, The Netherlands}{1986}{Broek}

\jref{cy}{Kin S.~Cheung and Sidney Yip}{\PRL}{65}{2804}{1990}{Cheung}
{Brittle-Ductile Transition in Intrinsic Fracture Behavior of Crystals}

\jref{vref1}{D.~R. Clarke and K.~T. Faber}{J.\ Phys.\ Chem.\ Solids }{48}
{1115}{1987}{Clarke}{Fracture of Ceramics and Glasses}

\jref{path4}{B.~Cotterell and J.~R.~Rice}{\IJF}{16}{155}{1980}{Cotterell}
{Slightly curved or kinked cracks}

\jref{bc}{W.~A.~Curtin and H.~Scher}{\PRL}{67}{2457}{1991}{Curtin}
{Analytic Model for Scaling of Breakdown}

\jref{bifur2}{John P.~Dempsey, Mao-Kuen Kuo, and Diane L.~Bentley}{Int.\ J.\
Solids Structures}{22}{333}{1986}{Dempsey}{Dynamic effects in mode III crack
bifurcation}

\jref{path3}{Norman A.~Fleck, John W.~Hutchinson and Zhigang Suo}
{Int.\ J.\ Solids Structures}{27}{1683}{1991}{Fleck}
{Crack path selection in a brittle adhesive layer}

\jref{fgms}{Jay Fineberg, Steven P.~Gross, M.~Marder, and Harry
L.~Swinney}{\PRL}{67}{457}{1991}{Fineberg}{Instability in Dynamic Fracture}

\bref{fr}{L.~B.~Freund}{{\bf Dynamic Fracture Mechanics},
Cambridge}{1990}{Freund}

\jref{GS}{R.~V.~Gol'dstein and R.~L.~Salganik}{\IJF}{10}{507}{1974}
{Gol'dstein}{Brittle fracture of solids with arbitrary cracks}

\aref{fact}{Fran\c{c}ois Hourlier, Hubert d'Hondt, Michel Truchon, and
 Andr\'e Pineau}{228}{{\bf Multiaxial Fatigue}, ASTM STP 853}{YEAR??}
{Hourlier}{Fatigue crack Path Behavior Under Polymodal Fatigue}

\jref{vref2}{J.~M. Huntley}{Proc.\ R.\ Soc.\ Lond.\ A}{430}{525}{1990}
{Huntley}{Crack growth in viscoelastic media: effect of specimen size}

\pref{path1}{Anthony R.~Ingraffea, Tulio N.~Bittencourt, and Jose Luiz
 Antunes O.~Sousa}{to appear in proc. of MECOM90: XI Congress
 Ibero-Americana Sobre Metodos Computacionais en Engenharia}{1990}{Ingraffea}
{Automatic fracture propagation for 2D finite element models}

\bref{kp}{Melvin F.~Kanninen and Carl H.~Popelar}{{\bf Advanced Fracture
Mechanics}, Oxford}{1985}{Kanninen}

\pref{kj}{Stephen A. Langer and Raymond E. Goldstein}{to be published; we
also thank Karsten Jacobsen for pointing out the particular appropriateness
of the term ``gauge'' for this problem}{1991}{Langer}{Dissipative Dynamics
of Closed Curves in Two Dimensions}

\jref{revx1}{Brian R. Lawn, David H. Roach, and Robb M. Thomson}{\JMS}{22}
{4036}{1987}{Lawn}{Thresholds and reversibility in brittle cracks:
an atomic surface force model}

\jref{revx2}{B.~R. Lawn and S.~Lathabai}{Mat.\ Forum}{11}{313}{1988}{Lawn}
{Surface forces and Fracture in Brittle Materials}

\pref{path3D}{J.~B.~Leblond}{preprint, presented at the International
Conference on Mixed Mode Fracture and Fatigue, Vienna}{1991}{Leblond}
{Crack kinking and curving in three-dimensional elastic solids---application
to the study of crack path stability in hydraulic fracturing}

\jref{vref3}{M.~Marder}{\PRL}{66}{2484}{1991}{Marder}
{New Dynamical Equation for Cracks}

\jref{pm}{Paul Meakin}{Science}{252 (12 April)}{226}{1991}{Meakin}{Models for
Material Failure and Deformation}

\jref{revt}{J.~R. Rice}{J.\ Mech.\ Phys.\ Solids}{26}{61}{1978}{Rice}
{Thermodynamics of the quasi-static growth of Griffith cracks}

\jref{path2}{Asher A.~Rubinstein}{\IJF}{47}{291}{1991}{Rubinstein}
{Mechanics of the crack path formation}

\bref{QFT}{Lewis H.~Ryder}{{\bf Quantum Field Theory} (see chapter 3),
Cambridge}{1985}{Ryder}

\jref{bifur1}{A.~Shukla, H.~Nigam, and H.~Zervas}{Engr.\ Fracture Mech.\ }
{36}{429}{1990}{Shukla}{Effect of stress field parameters on dynamic crack
branching}

\jref{revx3}{Kai-Tak Wan, Nicholas Aimard, S.~Lathabai, Roger G. Horn, and
Brian R. Lawn}{J.\ Mater.\ Res.\ }{5}{172}{1990}{Wan}
{Interfacial energy states fo moisture-exposed cracks in mica}

% FIGITEMS:
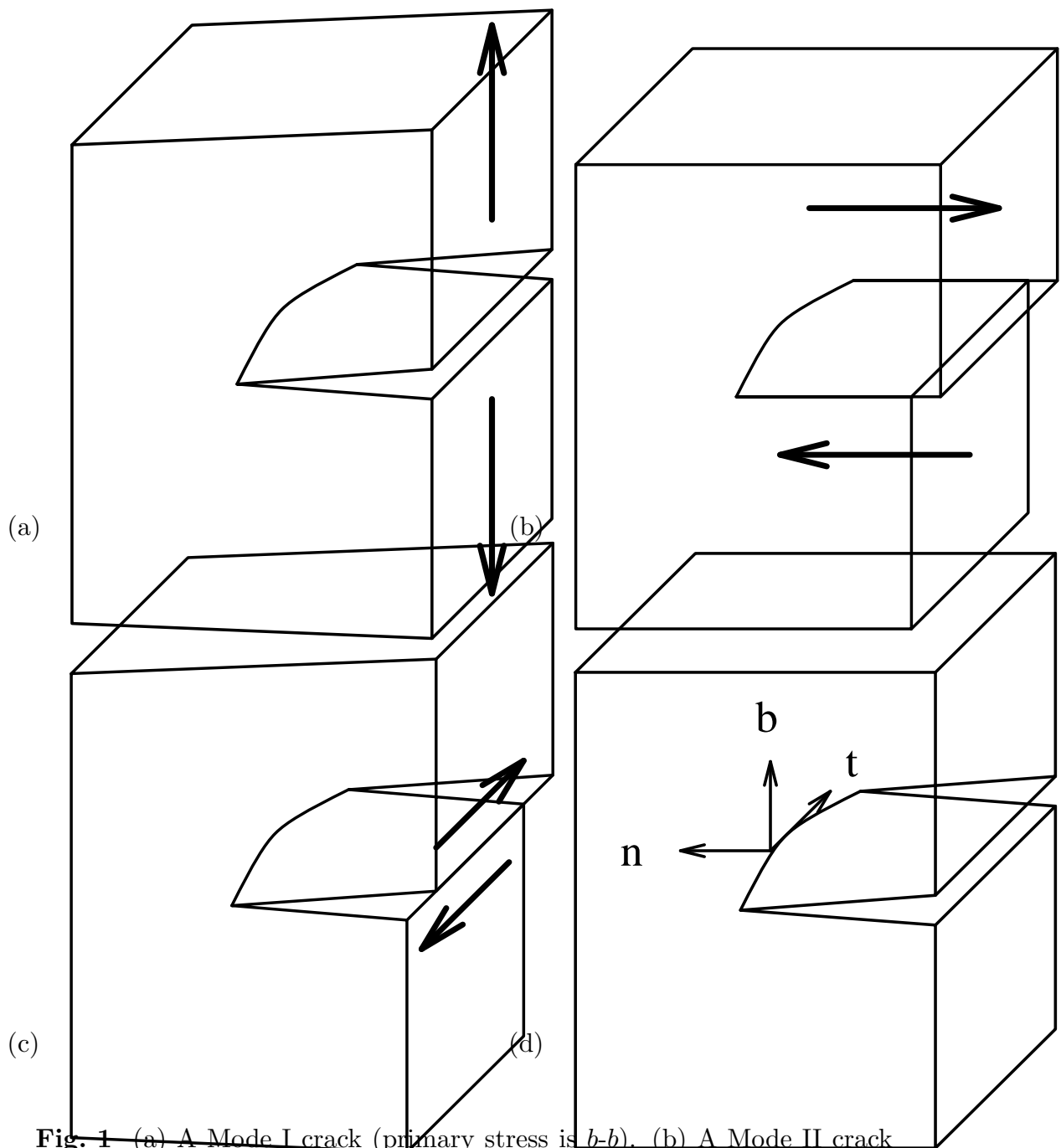
\figitem{fig1}{(a)~A Mode I crack (primary stress is $b$-$b$). (b)~A
Mode II crack (primary stress is $b$-$n$). (c)~A Mode III crack (primary
stress is $b$-$t$).
The local stress field around a crack can always be decomposed into
a linear sum of the three modes;  the decomposition can be
found from the modes' different symmetry properties (see appendix~B).
(d)~The vectors associated
with a point on the crack front: $\that$ is the tangent to the crack front
curve; $\nhat$, perpendicular to $\that$ and in the crack plane, is the
direction of crack growth; $\bhat\equiv\that\times\nhat$ is the normal to
the crack plane.}

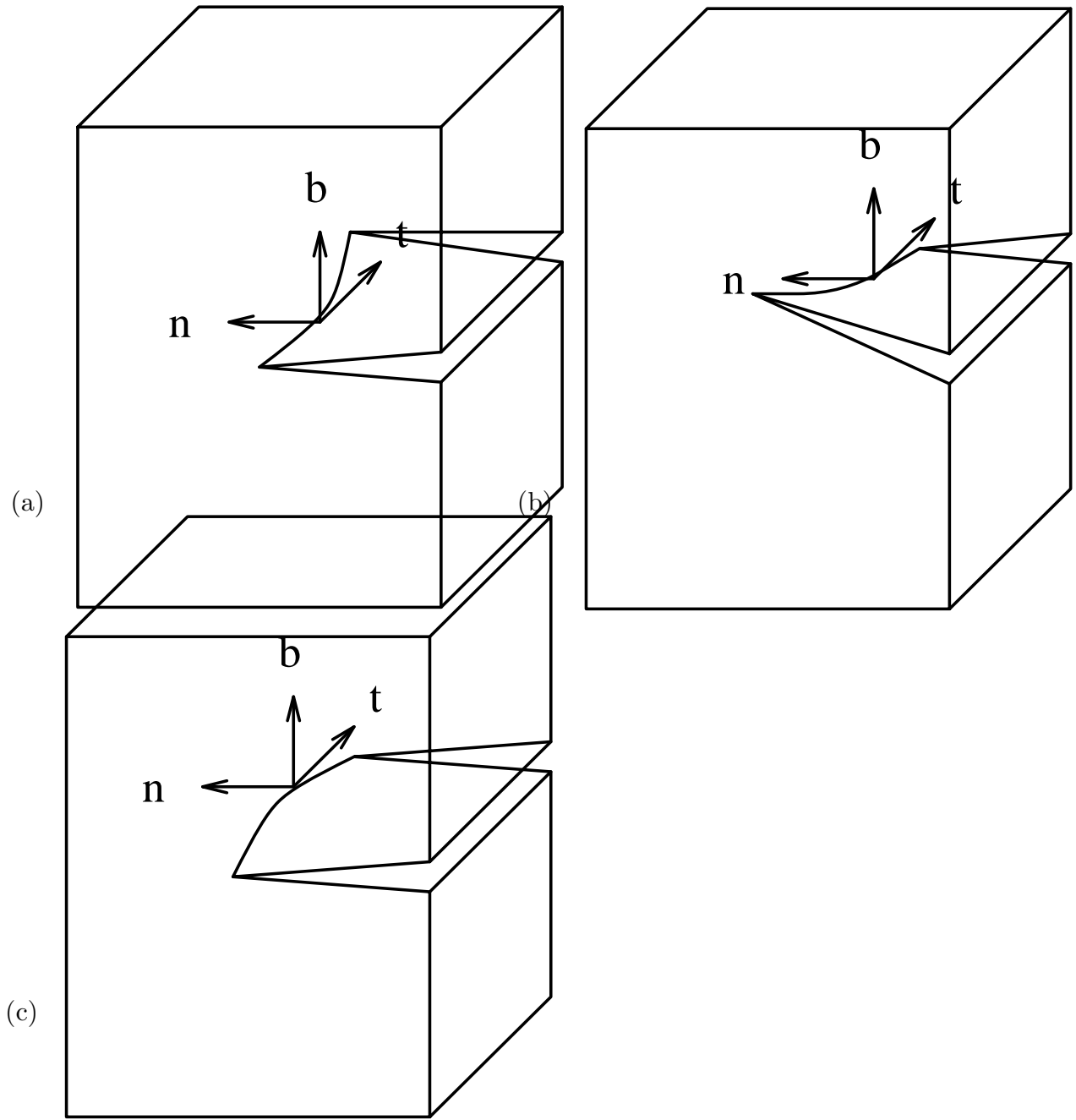
\figitem{fig2}{Transformations (a)~rotation about $\nhat$ and (b)~reflection
in the $n$-$t$ plane on (c)~the untransformed crack.  To make a physical crack
growth law, the crack velocity must be unchanged in both cases (if it
changed sign, the crack would heal); $\ddx{t}{\nhat(\lambda, t)} \cdot \bhat$
must change sign in both cases, since the direction of $\bhat$ is unaffected
by the transformation, while the {\it physical} normal to the crack plane
shifts to the opposite direction.}

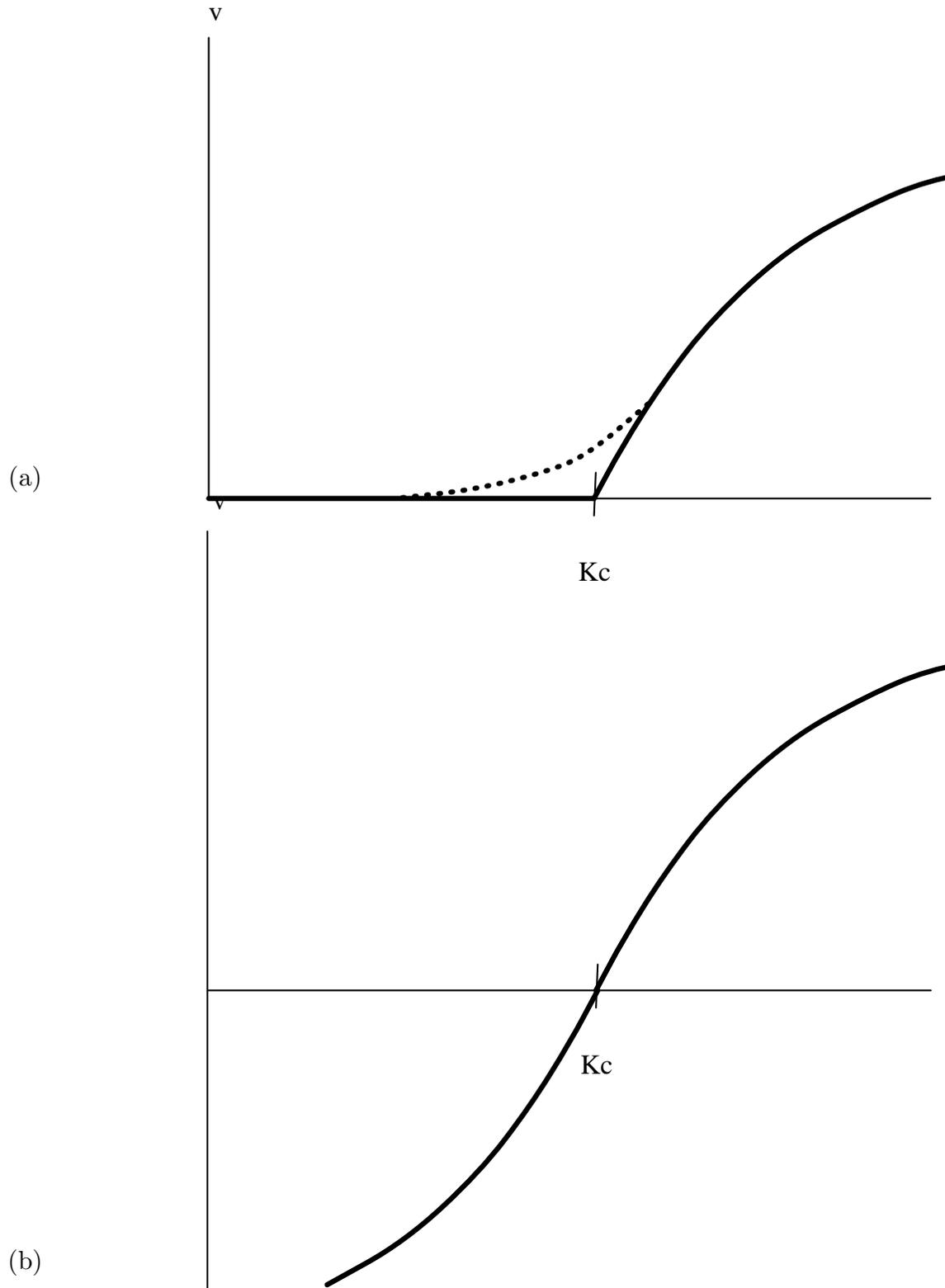
\figitem{fig3}{(a)~Form of the crack velocity $v$ as a function of stress
intensity factor $K$ in ordinary crack experiments, where the crack velocity
is zero below a critical value $K_c$ of the stress intensity factor, and
then has a sharp turn-on.  The
dotted line shows the behavior under fatigue, where the crack velocity
increases more gradually.
(b)~Form of the crack
velocity in very clean experiments, where crack healing can take place.}

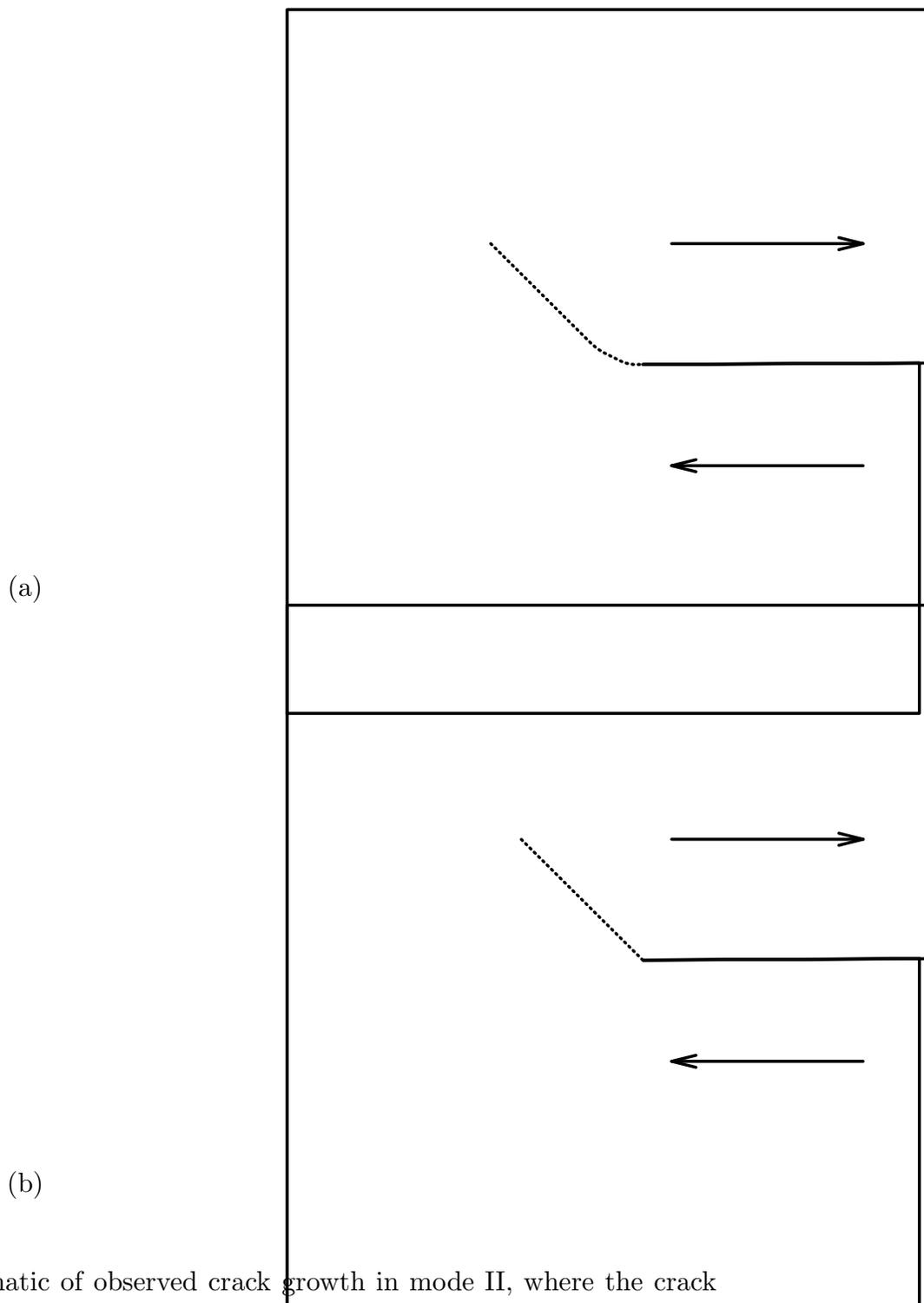
\figitem{fig4}{Schematic of observed crack growth in mode II, where the
crack curves so as to reduce the mode II stress, leaving only mode I
stress.  (a)~Our picture, where the crack curves gradually to the direction
where $\kII = 0$, on a length scale of $2 v \over f \kI$.  (b)~The
traditional picture, where there is a sharp kink to the direction where
$\kII = 0$.  Note that in the $f \rightarrow \infty$ limit, the two pictures
agree.}

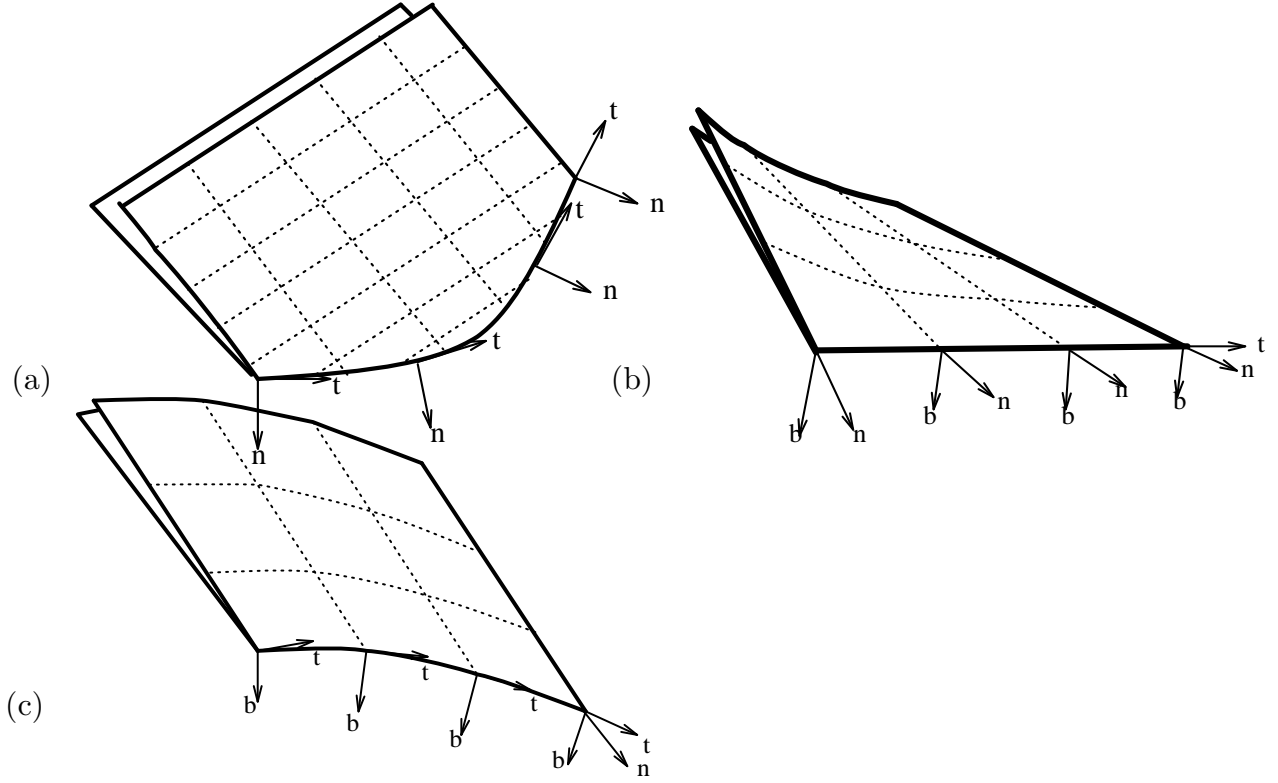
\figitem{fig5}{(a)~Top view of the surface of a planar crack with a curved
crack front, which gives a
non-zero value for $ \dadotb \nhat \that$, with both $ \dadotb \nhat \bhat$
and $ \dadotb \that \bhat$ zero (since
both $\nhat$ and $\that$ are always in the same plane, perpendicular to
$\bhat$).  Under
rotation of the crack about $\nhat$ at a point,
keeping the vectors fixed (symmetry operation~a), the value
of $ \dadotb \nhat \that$ stays the same; under reflection of the crack
in the $n$-$t$
plane at a point, again keeping the vectors fixed (symmetry operation~b),
it is also invariant.
(b)~A non-planar crack with non-zero $ \dadotb \nhat \bhat$, where
both $ \dadotb \nhat \that$ and $ \dadotb \that \bhat$ are zero (since $\bhat$
and $\nhat$ are always in the same plane, perpendicular to $\that$).
Under symmetry operation~a, $ \dadotb \nhat \bhat$ is invariant; it changes
sign under symmetry operation~b.
(c)~A non-planar crack with non-zero $ \dadotb \that \bhat$, where
both $ \dadotb \nhat \that$ and $ \dadotb \nhat \bhat$ are zero (since $\bhat$
and $\that$ are always in the same plane, perpendicular to $\nhat$).
$ \dadotb \that \bhat$  changes sign under symmetry operations a and b.
}

% TABITEMS:
\tabitem{t1}{Transformation properties of relevant variables and
their derivatives under two symmetry operations; see figure 5 and text.
Note that products of these quantities transform as the product of the
transformation properties
(e.~g.\ $(\kII \kIII)$ is $-$ under (a) and $+$ under (b)).}

%  TITLE  and ABSTRACT:
\skipline
\uppercase\expandafter{\centerline{\bf
Beyond the ``principle of local symmetry'':
}}
\uppercase\expandafter{\centerline{\bf
derivation of a general crack propagation law
}}
\skipline
{\begingroup\singlespace
 \centerline{Jennifer Hodgdon}
 \centerline{Laboratory of Atomic and Solid State Physics, Cornell
             University, Ithaca, NY 14853}
 \centerline{James P. Sethna\footnote{$^\ddagger$}{Permanent Address:
          Laboratory of Atomic and Solid State Physics, Cornell University,
          Ithaca, NY 14853}}
 \centerline{Laboratory of Applied Physics, Technical University of
             Denmark,}
 \centerline{DK-2800 Lyngby, DENMARK, and NORDITA,
             DK-2100 Copenhagen \O, DENMARK}

 \skipline
 \centerline{(January 30, 1992)}
 \skipline
 \centerline{\sl Submitted to the International Journal of Solids and
                 Structures}
 \skipline
\topicname{Abstract}

We derive  a general crack propagation law
for slow brittle cracking, in two and three dimensions, using symmetry,
gauge invariance, and gradient expansions.  Our derivation provides explicit
justification for the ``principle of local symmetry,'' which has been used
extensively to describe two dimensional crack growth, but goes beyond that
principle to describe three dimensional crack phenomena as well.  We also
find that there are new materials properties needed to describe the growth
of general cracks in three dimensions, besides the
fracture toughness and elastic constants previously used to describe cracking.
 \endgroup}

% TEXT:
\topicname{I. Introduction}

There are many aspects of the problem of crack growth
that have received a lot of attention recently. For instance,
there has been much interest in dynamic fracture\cite{fr} and the
accompanying crack bifurcation\cite{bifur1, bifur2} and other
instabilities.\pcite{fgms}  The transition between failure due
to percolation of a network of many small cracks and failure due
to a single dominating crack has also been explored,\pcite{bc} as well
as the transition between brittle and ductile cracking.\pcite{cy}
Pattern formation in multiple cracking\cite{pm} has also been of interest.
In light of all the interest in these rather complex phenomena of fracture,
it is somewhat surprising to find that little is known about the growth
laws for even slow-growing, simple (not multiple) three dimensional
cracks, though there has been some work done on calculating the
paths of cracks in two\cite{path1, path2, path3, path4}
and three\cite{path3D} dimensions,
and many measurements and calculations of
the crack velocity for simple two-dimensional
geometries.\pcite{vref1, vref2, vref3, bdl}

The problem of finding a growth law for cracks would seem to be of
fundamental interest; so,
in this paper, we apply the standard tools of theoretical
physics---gradient expansions, symmetry, and gauge symmetry---to find the
most general possible growth law for a three dimensional crack growing
slowly in a homogeneous, isotropic medium.
Since it is possible to make  precise numerical computations of the elastic
fields for arbitrary three-dimensional geometries,
with today's computers, in a matter of hours, we do not consider the
related problem of finding the stress state of the material containing the
crack, but consider it to be completely known.  We also compare the crack
growth law we derive here to previously derived and measured properties of
cracks in two and three dimensions.  In a second paper, we
will discuss the detailed behavior of cracks growing under this
law, using linear stability analysis as well as numerical
simulations; in a third paper we will discuss experiments designed
to measure material-dependent parameters appearing in the crack growth law.

\topicname{II. Simplifications}

We begin by simplifying the problem of crack propagation using
length and time scale considerations.
First, we smooth our crack problem over the length scale $\ell_s$
which characterizes the size of inhomogeneities and anisotropies in the
material containing the crack. (For example, in a glass,
$\ell_s$ is a few atomic sizes; in a polycrystal, it is
the grain size; in concrete, it is the size or distance
between the pebbles it contains.)  Although for a
single crystal, $\ell_s$ is as large as the body containing
the crack, for many situations of practical relevance, $\ell_s$ is much
smaller than the size of the body. In those cases, we can smooth
the crack problem over $\ell_s$ without losing much information, making
the crack a smooth surface, and the material containing the crack continuous,
homogeneous, and isotropic.

A second length scale in crack propagation problems arises because
every material has some stress above which it fails to have linear
elastic properties (e.~g.\ begin to flow, with plastic or viscous behavior;
emit dislocations; break bonds; or have a martensitic transformation).
For some materials, this stress is very low, and there is
no linear elastic regime at all.  For others, linear
elasticity is valid except very near the crack tip, where the stress
is much higher than in the bulk of the body.  For these materials,
there is a length
scale $\ell_{nl}$ which characterizes
the size of the non-linear process zone around the crack tip;
$\ell_{nl}$ can range from a few
angstroms in glass to tens of centimeters in concrete.
In this analysis, we consider only materials
for which $\ell_{nl}$ is small compared to the length of the
crack and the size of the body, so that the bulk of the
material can be considered linear elastic.
This work, then, describes materials
usually considered linear and brittle, as well as
materials exhibiting viscoelasticity, plasticity, and
martensitic transformation toughening, as long as the length scale for these
behaviors is sufficiently small. In principle, the
non-linear properties of such materials could be included in a
later version of this work
to extend its applicability to smaller length scales.

A third length scale relevant to crack propagation is associated with
the degree of translational invariance along the crack front.  For
many crack systems studied in the past, every plane perpendicular to
the crack front is equivalent, which means that the problems can be
considered two dimensional.  On the other hand, most practical crack
problems are not two dimensional, but instead have crack front
curvature or
stresses which vary along the crack front.  If this is the case, then
there is a length scale, which we call the
dimensional crossover length $\ell_{dc}$,
above which the problem is three dimensional;
$\ell_{dc}$ is either one of the geometric lengths associated with the crack
geometry (such as
the radius of curvature of the crack
front), or is associated with the stress gradient:
$\ell_{dc} \approx {\sigma \over \del \sigma}$.
For this work, we assume that $\ell_{dc}$
is large, though in general not as
large as the size of the body containing the crack, and we expand
in powers of quantities which are inversely proportional to $\ell_{dc}$
(i.~e.\ gradients).

Finally, we also simplify the crack propagation problem by considering
cracks which are growing slowly enough that inertial and relativistic
(relative to the {\it sound,} not light, velocity)
effects are unimportant.  Some of these effects could be included in
future work, but the present analysis suffices for cracks which
arrest after growing a certain distance, such as when a wedge is
driven into a crack; cracks which grow at a constant speed,
such as under constant displacement loading; and cracks which
may eventually speed up, but which are currently growing slowly, as
in the cases of fatigue cracks, sub-critical cracking, and the first
stages of growth under constant force loading.

\topicname{III. Relevant variables}

Although the knowledge of length scales from the previous section has
simplified our problem to a nearly two dimensional, isotropic,
homogeneous, linear-elastic, continuous bulk medium with a smooth
crack, at first glance it appears that there are still many variables
which could influence the propagation of the crack; for example, the
load on the surface of the body, type of material, temperature,
ambient atmosphere, and the stress and fracture history.  However,
we are concerned here with the propagation of a crack given the
elastic fields in the body, not the precise conditions that produced
those fields.  Also, many variables, such as stress history and
temperature, can be included implicitly in the materials constants,
which we also assume are known.  This means that for this work,
the variables of relevance are the elastic fields near the
crack tip, materials constants,
and the current configuration of the crack.

Now, it is well known that for cracks in
linear elastic media, the stress field near the tip of the crack---which is
the only area we expect to influence crack growth---obeys a
$\sim {1\over\sqrt r}$ power law in the distance $r$
from the crack tip in the plane.  There are three modes of cracking (see
figure 1a, b, and c); each has a characteristic known angular
dependence, which can be written\cite{kp}
$$\sigma_{ij}(r,\theta) =
  { K_\alpha \over \sqrt{2 \pi r}} f^\alpha_{ij}(\theta), \putn{eq1}$$
where $\theta$ is the angle made with $\nhat$, the direction of cracking
(see figure 1d), and the three $K_\alpha$ are the
mode I, II, and III Stress Intensity Factors (SIFs), numbers characterizing the
strength of the stress singularity for each cracking mode.  There are
also similar expressions for the displacement field near the crack
(see reference \refnumber{kp}).
 From equation \equn{eq1}, we can see that the relevant information
from the elastic
field for a two dimensional crack propagation problem, instead of being given
by a displacement vector at every point in the plane,
is reduced to just three numbers, the SIFs.   In weakly
three dimensional problems, each non-equivalent plane has three SIFs
which characterize the stress, so that the relevant
information from the elastic field is given by
three numbers at each point on the crack front,
instead of a displacement vector at every point in the body.

\pageinsert{
\hbox to 6.5 truein{
(a)
\psfig{figure=mode1.ps,width=3.0truein}
(b)
\psfig{figure=mode2.ps,width=3.0truein}
 }
\hbox to 6.5truein{
(c)
\psfig{figure=mode3.ps,width=3.0truein}
(d)
\psfig{figure=vecs.ps,width=3.0truein}
}
\figcap{fig1}
}
\endinsert

Now, the configuration of the crack could be specified by giving equations
for the two surfaces of the crack, but it can also be reduced to a
smaller amount of information.  This comes about because
part of the information
on the crack geometry is contained in
the elastic fields, and the part of the
crack surface far from the crack tip cannot influence the growth of the
crack beyond affecting the elastic fields.  So, the most relevant information
about the crack geometry, aside from the
elastic field, is given by the crack front curve,
$\xvec(\lambda)$, and the vector $\nhat(\lambda)$
giving the current direction of
crack growth, where $\lambda$ parameterizes the crack front curve.
In figure 1d we show the three unit
vectors associated with this description:
$\that(\lambda) \equiv \ddx{s}{\xvec}$, the tangent to the curve, with $s$
the arc length;
$\nhat(\lambda)$, the direction of
crack growth, perpendicular to $\that(\lambda)$
(so that $\that(\lambda)$ and $\nhat(\lambda)$
define the crack plane at point $\lambda$), and
$\bhat(\lambda) \equiv \that(\lambda) \times \nhat(\lambda)$,
the normal to the local crack plane.
Note that since the material containing the crack is
isotropic, the coordinate system defined by these unit vectors is the
only one physically relevant to crack growth, and all other quantities
(such as the SIFs) are understood to be defined in this coordinate system
(see figure 1).
Also, in two dimensional cracks, $\that$, $\nhat$, and $\bhat$ are all constant
along the crack front.

\topicname{IV. The crack growth law}

Now we are ready to derive a crack growth law in our relevant variables:
materials constants, the SIFs $K_\alpha(\lambda)$,
and the unit vectors $\nhat(\lambda)$,
$\that(\lambda)$, and $\bhat(\lambda)$.
That is, we are ready to derive an expression for
the time evolution of  the crack front curve $\xvec(\lambda)$ as an expansion
in $\dds \equiv \left( \ddx \lambda s \right)^{-1}\dop \lambda$,
the gradient along the crack curve, of the relevant variables.
(In two dimensions, $\dds = 0$.)
Noting that the crack surface is smooth for all time, we know that the
time derivative of $\xvec$ must lie in the crack plane, so we can write
$$\ddx{t}{\xvec(\lambda,t)} = v(\lambda,t) \nhat(\lambda,t) +
          w(\lambda,t) \that(\lambda,t), \putn{eq2}$$
where $v$ is the crack velocity, and $w$ is a non-physical function
which can be chosen freely to determine how
$\lambda$ parameterizes the physical crack surface. That is, a
particular choice of $w$ determines a gauge for $\lambda$
(see appendix~A);  in two dimensions, we take $w = 0$.

To determine the growth of the physical crack, then, we need
to find $v(\lambda,t)$ and $\nhat(\lambda,t)$ in equation \equn{eq2}, in
terms of the relevant variables. Noting that $\nhat(\lambda,0)$ is given,
this means that to find $\nhat(\lambda,t)$, we need to find the time
derivative $\ddx t \nhat$.
Now, since $\nhat$ is a unit vector, its time derivative must be
perpendicular to itself; also, by definition $\nhat$ is perpendicular to
$\that \equiv \ddx{s}{\xvec}$.  This gives us a constraint on the equation
of motion for $\nhat$, obtained by setting $\dop{t} (\nhat \cdot \that) = 0:$
$$\ddx{t}{\nhat} \cdot \that = - \ddx{s}{v} -
   w \ddx{s}{\that} \cdot \nhat. \putn{eq3}$$
Note that in two dimensions the right hand side of this equation vanishes,
since $\dds = 0$.

Another constraint on $\ddx{t}{\nhat}$, and one on
the crack velocity $v$ are obtained from
symmetry.
We consider symmetry operations, centered at some point on the crack
front, which leave the unit vectors at that
point fixed and reflect or rotate the material, preserving the
physical properties that $\that$ is the tangent to the crack front curve
and $\nhat$ is the direction of cracking at that point.  (This is equivalent
to leaving the material fixed and transforming the coordinates.)
There are two such independent operations:
(a)~\deg{180} rotation about $\nhat$, and (b)~reflection in
the $n$-$t$ plane (the crack plane) (see figure~2).
Under both of these operations, the
physical law for $v$ must remain
unchanged; that for $\ddx{t}{\nhat} \cdot \bhat$ must change
sign under both operations (see figure~2).

\pageinsert{
\hbox to 6.5 truein{
(a)
\psfig{figure=rotat.ps,width=3.0truein}
(b)
\psfig{figure=reflect.ps,width=3.0truein}
 }
\hbox to 3.25truein{
(c)
\psfig{figure=notrans.ps,width=3.0truein}
}
\figcap{fig2}
}
\endinsert

Let us now examine the case of a crack in two dimensions, where the only
non-vector quantities we can form are combinations of the SIFs and
material constants.  Under symmetry operation (a), the material constants,
$\kI$, and $\kIII$ remain the same, while $\kII$ changes sign; under operation
(b), both $\kII$ and $\kIII$ change sign, and everything else remains the same
(see appendix~B).  Since the three SIFs have different transformation
properties, and only $\kII$ transforms like $\ddx{t}\nhat \cdot \bhat$,
we see that the two-dimensional crack growth law must have the form:
$$ \eqalign{
 \ddx{t}{\xvec} &= v \nhat \cr
 \ddx{t}{\nhat} &= - f \kII \bhat, \cr } \putn{eq4}$$
where both $v$ and $f$ are functions of materials constants,
$\kI$, $\kII^2$, and $\kIII^2$.
(The dependence of $f$ and $v$ on $\kII$ and $\kIII$ must
be quadratic to insure invariance under both symmetry operations.)
The minus sign makes $f > 0$ correspond
to the observed direction of crack growth under mode II loading;
this is discussed in the next section.

In three dimensions, the gradient $\dds$ is not
strictly zero, and there are therefore non-vector quantities besides the SIFs
and materials constants which can be
formed from the relevant variables.  Up to first order in $\dds$,
these are listed, along with their transformation properties,
in table~1 of appendix~B.  From
the transformation properties of these quantities,
and from the discussion in appendix~A on gauge invariance,
we can see that the physical crack growth law, to
first order in $\dds$,
has the general three-dimensional form:
$$ \eqalign{
 \ddx{t}{\xvec} =& v \nhat + w \that \cr
 \ddx{t}{\nhat} =& - \left[ \ddx{s}{v} +
   w \ddx{s}{\that} \cdot \nhat \right] \that + \cr
&  \left[ - f \kII + g_{\rm I} \kIII \ddx{s}{\kI} +
   g_{\rm II}\kII \kIII \ddx{s}{\kII} + g_{\rm III}\ddx{s}{\kIII} + \right. \cr
&\qquad \left.  h_{tb} \dadotb{\that}{\bhat} +
   h_{nt} \kII \dadotb{\nhat}{\that} +
 (h_{nb}\kII\kIII + w)  \dadotb{\nhat}{\bhat} \right] \bhat ,\cr } \putn{eq5}$$
where the $f$, $g_\alpha$, and $h_{ij}$ are functions of materials constants,
$\kI$,  $\kII^2$, and $\kIII^2$; and $v$ is a function of materials constants,
$\kI$,  $\kII^2$, $\kIII^2$, $(\kII \kIII \ddx s \kI)$, $(\kIII \ddx s \kII)$,
$(\kII \ddx s \kIII)$, $(\kII  \dadotb{\that}{\bhat})$,
$(\dadotb{\nhat}{\that})$, and $(\kIII \dadotb{\nhat}{\bhat})$, since
these are all the quantities up to first order in $\dds$ which are invariant
under both  symmetry operations.

\topicname{V. The undetermined functions in the crack growth laws}

Now, in equations \equn{eq4} and \equn{eq5}
we have general forms for the crack growth law for two and
three dimensional cracks.  These equations contain many unspecified functions,
which must be determined from considerations  other than
the symmetry considerations we used to find the general forms.
Perhaps the most important of these functions is the crack growth velocity,
$v$, a function of materials constants, $\kI$, $\kII^2$, and $\kIII^2$
in two dimensions.  This function has been measured for
mode I cracks\cite{vref1} and usually has the form shown in
figure 3a, with zero velocity for $\kI$ below some value ${\kI}_c$,
which depends on the material,
and a monotonically increasing velocity above ${\kI}_c$.
This schematic form for the velocity has also been found in a theoretical
calculation for a viscoelastic system.\pcite{bdl}
However, both in theory\cite{revt, revx1, revx2, revx3} and in very
clean experimental systems,\pcite{revx1, revx2, revx3} the crack velocity has
the form in figure 3b, with
a negative velocity (crack healing) when $\kI < {\kI}_c$.
This means that the crack
velocity is a continuous function which passes through zero at ${\kI}_c$,
so that for SIFs near ${\kI}_c$, (i.~e.\ small
crack velocities), we can expand $v$ as
$v(\kI) \approx v_0 { \kI - {\kI}_c \over {\kI}_c}$,  with $v_0$ a
material dependent constant.  For modes
II and III cracks, as well as mixed mode cracks, since the elastic energy
released per unit area of crack surface (the ``energy release rate'') is
proportional to $(\kI^2 + \kII^2 + {\kIII^2 \over (1-\nu)})$, where $\nu$ is
Poisson's ratio,\pcite{eefm} we expect that a crack
velocity function valid for  all modes of cracking can be expanded as:
$$v(\kI, \kII, \kIII) \approx v_0 {K - K_c \over K_c}, \putn{veleq}$$
where $K \equiv (\kI^2 + \kII^2 + {\kIII^2 \over (1-\nu)})^\half$.
Also, note that for fatigue cracking, where our growth laws must still
hold (on time scales long compared to the load cycle),
the crack velocity generally does not
go to zero sharply at $K_c$, but has a more gradual turn-on
behavior\cite{eefm} (see figure~3a).

\pageinsert{
\hbox to 6.5 truein{
(a)
\psfig{figure=vela.ps,width=5.0truein}}
\hbox to 6.5 truein{
(b)
\psfig{figure=velb.ps,width=5.0truein}}
\figcap{fig3}
}
\endinsert

Now, we saw in the previous section that the crack velocity in three
dimensions can also depend on gradient quantities, besides the SIFs;  the
dependence of the crack velocity on these quantities
has not been measured, to our knowledge.
However,
there is no reason to suppose that the dependence on these
quantities has special behavior (e.~g.\ zero-crossing or
very strong dependence on SIFs) near $K_c$,
the value of the SIF where the crack velocity becomes positive. So,
for small velocities, where $K \approx K_c$,
we can approximate the dependence of the crack velocity on these quantities
by a constant function (i.~e.\ its value at $K_c$).

Similarly, we expect that the seven
functions $f$, $g_\alpha$, and $h_{ij}$ in equations \equn{eq4} and
\equn{eq5}, which are allowed by symmetry to be functions  of
materials constants, $\kI$,
$\kII^2$, and $\kIII^2$, can be approximated as constants when the
velocity is small.  This means that to find the material-specific form of
the crack growth law, for small velocities, it is a reasonable
approximation to measure only the linearized
dependence of the crack velocity on $K$, and the constant parts of
$f$, $g_\alpha$, and $h_{ij}$.

\topicname{VI. Predictions of the crack growth laws}

Let us now
examine the two dimensional crack growth law, equation
\equn{eq4}:
$$ \eqalign{
 \ddx{t}{\xvec} &= v \nhat \cr
 \ddx{t}{\nhat} &= - f \kII \bhat. \cr } \putn{eq4}$$
When $\kII = 0$, this equation says that the crack
grows in a straight line (since $\ddx{t}{\nhat} = 0$),
in agreement with the ``principle of local symmetry''\cite{GS} generally used
to predict crack growth in two dimensions.  However, the principle of local
symmetry also
says that $\kII = 0$ is maintained at all times by the propagating
crack---in effect, that the crack curves in such a way as to keep
$\kII=0$. Our law, in contrast, says that it is only a non-zero $\kII$ which
can make the crack curve, but that (with $f>0$) the crack curves in such a
way as to make $\kII$ smaller (see figure 4).

\pageinsert{
\hbox to 6.5 truein{
(a)
\psfig{figure=mode2growtha.ps,width=4.0truein}}
\hbox to 6.5 truein{
(b)
\psfig{figure=mode2growthb.ps,width=4.0truein}}
\figcap{fig4}
}
\endinsert

Now, we can resolve the differences between the principle of local symmetry
and our crack propagation law by writing the crack velocity $v$ as the
time derivative of the crack length $\dadt$,  writing
$\dop t = \dadt \dop a$,  and by writing $\nhat$ and $\bhat$ in terms of the
angle $\theta$ that $\nhat$ makes with the $x$-axis.
With these changes, equation \equn{eq4} becomes:
$$ \eqalign{
 \ddx{a}{\xvec} &= (\cos\theta, \sin\theta) \cr
 \ddx{a}{\theta} &= - \left({f\over v}\right) \kII . \cr } \putn{eq6}$$
In principle, $f$, $v$, and $\kII$ are functions of $\xvec$ and $\theta$.
However, in the case of a small amount of growth at the end of a
long crack, we expect $f$ and $v$ to be nearly
constant as the crack grows, since the SIFs only change by a small
amount during the growth  (see appendix~C.)
Also, when $\theta$ differs from the angle that makes $\kII = 0$ by only a
small amount $\Delta \theta(x)$, we can approximate $\kII$ as
\def\equat9{
$$\kII(x) = \kI(0) {\Delta\theta(x) \over 2}
   \left[1+{\cal O}\left({x\over a}\right)\right], \putn{eq9}$$}
\equat9
where $a$ is the length of the original long crack (see appendix~C).
Using equation \equn{eq9} in equation \equn{eq6}, we see that
$$\ddx{a}{\Delta\theta} = - \left({f \kI(0)\over 2 v}\right) \Delta\theta,
  \putn{eq10}$$
which we can immediately solve, taking $f$ and $v$ constant, to find that
$$\Delta\theta(a) = e^{- {f \kI(0) a \over 2 v} }. \putn{eq11}$$
That is, if we start a crack with a small deviation from the direction
predicted by the principle of local symmetry, then our crack propagation law
says that the deviation decays with a characteristic distance of
${2 v \over f \kI}$. This length scale must be very small, about the size of
the non-linear process zone or the smoothing length, because these
microscopic lengths are the only length
scales that appear in two-dimensional crack problems (see section II).
In the limit that the length scale is zero, or $f \rightarrow \infty$,
we can now see that our crack
propagation law for two dimensions agrees with the principle of local
symmetry.

Now, let us move to consideration of equation \equn{eq5},
$$ \eqalign{
 \ddx{t}{\xvec} =& v \nhat + w \that \cr
 \ddx{t}{\nhat} =& - \left[ \ddx{s}{v} +
   w \ddx{s}{\that} \cdot \nhat \right] \that + \cr
&  \left[ - f \kII + g_{\rm I} \kIII \ddx{s}{\kI} +
   g_{\rm II}\kII \kIII \ddx{s}{\kII} + g_{\rm III}\ddx{s}{\kIII} + \right. \cr
&\qquad \left.  h_{tb} \dadotb{\that}{\bhat} +
   h_{nt} \kII \dadotb{\nhat}{\that} +
 (h_{nb}\kII\kIII + w)  \dadotb{\nhat}{\bhat} \right] \bhat ,\cr } \putn{eq5}$$
our crack propagation law for three dimensions.  First, note that the
principle of local symmetry term, $ - f \kII$, appears in this law, just as
in two dimensions, and we do not expect $f$ to be different here from its
value in two dimensions (that is, $f\over v \kI$ is the inverse of a
microscopic length).  In contrast to $f$, the other functions appearing in
equation \equn{eq5} do not contain any length scales---both
$g_\alpha \over v \kI$ and $h_{ij} \over v \kI$ are dimensionless.  The
length scales in these terms come from the gradient $\dds$; if the
dimensionless forms of $g$ and $h$
are of order 1, these terms act over the length scales of the
gradients.  As noted in the previous section, we do not know $g_\alpha$ or
$h_{ij}$ from symmetry principles; simple experiments to measure the sign
and magnitude of these material-dependent functions, and their physical
interpretations, will be discussed in a separate paper.

Our crack growth law, since it contains terms besides the principle of local
symmetry term, predicts three-dimensional behavior beyond the scope of the
principle of local symmetry; one example of this is the so-called "factory
roof" structure seen in mode III cracks.\pcite{fact} We will explore this
and other predictions of our crack propagation law more fully in a separate
paper, and also discuss the stability of straight cracks to wavy
perturbations.

\topicname{VII. Conclusions}

We have seen that from symmetry principles, we can derive a crack
growth law for both two and three dimensional geometries which agrees
with the principle of local symmetry\cite{GS} in the limit that
the microscopic length scales in the crack problem are truly zero.
Our law also predicts behavior seen in three dimensions that is not
predicted by the principle of local symmetry; more
detailed predictions of our crack growth law in three dimensions
will be explored in a separate paper.
The laws we derived, equations \equn{eq4} and \equn{eq5}, contain several
functions, such as the crack velocity,  which are not determined by symmetry
alone and must be measured in
controlled cracking experiments or atomic simulations; we will discuss
experiments designed to measure the sign and order of magnitude of
these functions, and their physical interpretations, in a third paper.

By using symmetry principles, separation of length scales, gauge invariance,
and gradient
expansions, then, we have derived effective, macroscopic equations governing
the growth of cracks in three dimensions.  We have course-grained the
problem so that microscopic details---such as atomic bond breaking,
crystalline grain morphology, deformation near the crack tip in response to
strain, and surface effects---are on such small length scales that they
cannot affect the macroscopic crack growth.  Understanding the microscopic
origins of our effective growth equations, and describing crack growth on
very small length scales, where our crack growth law is not valid,
will demand calculations that include these microscopic details.

\topicname{Acknowledgments}

We acknowledge the support of (US) DOE Grant \#DE-FG02-88-ER45364;
JH was funded by an NSF Graduate Fellowship and a
DOEd National Needs in Materials Physics fellowship.
We would like to thank P.~Wash Wawrzynek, Dave Potyondy, and Tony Ingraffea
for introducing us to the problem of crack propagation, for many
discussions about the subject, and for allowing us to use their
two-dimensional finite element analysis code (FRANC);
Jim Krumhansl for his continuing support and critical evaluation of this
research; and  Andy Ruina, Robb Thomson,  and Jim Rice for
their comments on earlier versions of this work.
JPS would like to thank Stephen Langer for introducing him to
the gauge invariance of curves.
We would also like to thank the
Technical University of Denmark and NORDITA for support and hospitality.

\topicname{Appendix A: Gauge symmetry and cracks}

There are many cases where the natural mathematical description of a problem
introduces fictitious degrees of freedom with no physical relevance.  The
most well-known example is in electromagnetism, where all physical quantities
are unchanged when the gradient of an arbitrary function $\chi(\vec r)$ is
added to the vector potential $\vec A$. The transformation
$\vec A \rightarrow \vec A + \del \chi$ is an example of a
``gauge transformation;''
the invariance of physical quantities under such a transformation is called
``gauge invariance.''  The strong and weak forces of particle physics also
have gauge transformations associated with them.\pcite{QFT}
In general relativity, the choice of coordinate system for space-time is
arbitrary; this gauge invariance can be used
to derive momentum and energy conservation.\pcite{QFT}
Another use of gauge symmetry is in site-disorder spin
glasses,\pcite{BY} where a gauge transformation is used
to show that the certain forms of disorder do not result in spin glass
properties, but in ferromagnetic behavior.

The term ``gauge''  is particularly appropriate for the gauge symmetry
of our problem,\pcite{kj}  where the parameterization $\lambda$ of
the crack front curve $\xvec(\lambda, t)$ is arbitrary: how one
``gauges'' (measures) the points along the curve cannot affect the
growth of the crack.  There are two different
types of gauge symmetry for cracks.  The first type is
the freedom to change the
parameterization at any one time, which we call the ``one-time gauge
symmetry.''
The second is the
freedom to choose how the parameterization at some time is related,
through the growth equation, to the parameterization at a later time; we
call this the ``time-dependent gauge symmetry.''
Crack growth laws
must satisfy both gauge symmetries; that is, neither
the one-time nor the time-dependent gauge transformation can change the {\it
physical} crack growth equation.

A crack growth equation that satisfies the one-time gauge symmetry
must not have any direct dependence on the value of $\lambda$ at a
point on the crack front curve, but only depend on physical quantities
evaluated at that value of $\lambda$. Also, derivatives along the
crack front cannot enter the growth equation as $\dop{\lambda}$, but
must instead be in terms of the arc length $s$, because $\dop s$ is
gauge invariant.  Since we have written our crack growth law
in terms of $\dds$, and have not included explicit $\lambda$-dependence,
it does satisfy the one-time gauge symmetry.

The time-dependent gauge symmetry is slightly more complicated.  An
equation for $\ddx{t}{\xvec(\lambda,t)}$, which is how we have chosen
to write the crack growth equation, tells us how a point with
parameter value $\lambda$ evolves in time.  This means that the time
evolution of the parameterization is implicit in our formulation, and
the time-dependent gauge transformation
changes the growth equation (unlike the one-time gauge transformation).
This change happens in a well-defined way:
if we have some crack growth law in terms of a parameter $\lambda$,
$$\eqalign{
 \ddx{t}{\xvec(\lambda,t)} &= A \nhat + B \that \cr
 \ddx{t}{\nhat(\lambda,t)} &= C \that + D \bhat, \cr} \putn{A1}$$
where the right hand sides are implicit functions of $\lambda$,
when we introduce a time-dependent gauge transformation to a new parameter
$\mu(\lambda,t)$, then the crack growth law becomes
$$\eqalign{
 \ddx{t}{\xvec(\mu,t)} &= A \nhat + B \that  + \ddx{\mu}{\xvec}\ddx{t}{\mu}\cr
 \ddx{t}{\nhat(\mu,t)} &= C \that + D \bhat + \ddx{\mu}{\nhat}\ddx{t}{\mu}\cr}
        \putn{A2}$$
with the right hand sides now implicit functions of $\mu$.
Writing $\dop{\mu} =  \ddx \mu s \dop s$
(with $s$ the arc length), defining a new function
$w \equiv \ddx \mu s \ddx t \mu$, and using the
definition of $\that \equiv \ddx s \xvec$, we can write this as
$$\eqalign{
 \ddx{t}{\xvec(\mu,t)} &= A \nhat + (B + w) \that \cr
 \ddx{t}{\nhat(\mu,t)} &= C \that + D \bhat + w \ddx s \nhat \cr
                       &= (C + w \dadotb \nhat \that) \that +
 			  (D + w \dadotb \nhat \bhat) \bhat. \cr}
        \putn{A3}$$

There are three particular time-dependent
gauges that we have found to be of special use in
our study of crack growth.  First, there is the ``reference gauge,'' where
$B + w$ in equation \equn{A3} is zero.  In this case, curves of constant
parameter value $\lambda_n$ are the integral curves of $\nhat$, and the
growth equation can be written:
$$\eqalign{
 \ddx{t}{\xvec(\lambda_n,t)} &= v \nhat \cr
 \ddx{t}{\nhat(\lambda_n,t)} &=  -\ddx s v \that + D \bhat. \cr} \putn{A4}$$
Here the $\ddx s v \that$ term comes from the requirement that
$\nhat \cdot \that = 0$ be preserved at all times; the function $D$ is
free, as far as gauge symmetry is concerned.  This is the only
possible form of the growth law when $\ddx t \xvec$ is along $\nhat$,
and we use this as our reference gauge for discussing other time-dependent
gauges.  In fact, comparing equation \equn{A4} to equations \equn{A1}  and
\equn{A3}, we see that the growth law in a general time-dependent
gauge can be written as:
$$\eqalign{
 \ddx{t}{\xvec(\mu,t)} &= v \nhat + w \that \cr
 \ddx{t}{\nhat(\mu,t)} &=  (-\ddx s v  + w \dadotb \nhat \that) \that
        + (D +  w \dadotb \nhat \bhat) \bhat, \cr} \putn{A5}$$
with $w = \ddx \mu s \ddx t \mu$ as above.  Note that the function $w$
characterizes the time-dependent gauge used in the crack growth law, while
$D$ and $v$ are the physical functions describing the crack growth.

Another particular time-dependent
gauge of interest is the ``arc length gauge,'' where the parameter
$\lambda$ is always equal to the arc length $s$
along the crack front curve, measured from some starting point (such as
the edge of the body).  Since the arc length along the crack front is
$s(\lambda) = \int_{\lambda_0}^{\lambda} ||\ddx \lambda \xvec || d \lambda$,
if we begin in the arc length gauge
(by making a one-time gauge transformation), then we can remain in that
gauge by choosing
$$w(\lambda_s) =
     \int_{\lambda_0}^{\lambda_s} v(\lambda') \ddx {\lambda'} \nhat \cdot
     \that(\lambda') d \lambda' \putn{A6}$$
in equation \equn{A5}.  Although arc length is physically the most natural
parameterization for a curve, the arc length gauge is not usually very
convenient, as $w(\lambda_s)$ is a non-local function---as the crack
grows, if the arc length of a section near some point $\lambda_1$ stretches (or
shrinks),
$\lambda_s$ must shift upwards (downwards) for all points with $\lambda_s >
\lambda_1$.

A third time-dependent gauge, the ``$z$ gauge,''
is useful for cracks which have crack fronts
which always point nearly along some axis, which we take to be the $z$-axis.
In this case, it is natural to use a gauge where the parameter $\lambda$
of the crack front is  the $z$-coordinate (as long as the crack
front $\xvec(\lambda_z)$ is a single-valued function).  To achieve this,
the crack growth equation for $\xvec$ must have zero $z$-component;
that is, we must choose
$$w(\lambda_z) = - v(\lambda_z) {\nhat_z \over \that_z}. \putn{A7}$$
Similar ideas can be used to make special gauges for cracks
which are nearly circular, parameterizing with the angle $\theta$ from
the $x$-axis, for instance, and for other common crack geometries.

\topicname{Appendix B:  Symmetry operations on the relevant variables}

In this appendix, we examine the two symmetry operations
(a)~\deg{180} rotation of the material about $\nhat$ at some
point $\lambda_0$, while keeping the coordinates fixed, and (b)~reflection
of the material in the $n$-$t$ plane (the crack plane) at $\lambda_0$,
while keeping the coordinates fixed (see figure 2),
as applied to the relevant variables for crack growth and their derivatives.
Note that only the signs, and not
the magnitudes, of the variables and derivatives can change under
these two operations.  Also note that the signs of the SIFs are
defined in terms of the three unit vectors, and that the
gradient operator $\dds$ along the crack front curve
goes in the direction of $\that$.

Now, $\kI$ transforms like the $\sigma_{bb}$ stress component, $\kII$ like
$\sigma_{nb}$, and $\kIII$ like $\sigma_{bt}$ (see figure 1).
Symmetry operation (a) takes material at $(x_n, x_b, x_t)$, in terms
of the coordinates with axes $\nhat$, $\bhat$, and $\that$ and origin
at $\lambda_0$, to
$(x_n, -x_b, -x_t)$; symmetry operation (b) takes material at $(x_n, x_b, x_t)$
to $(x_n, -x_b, x_t)$.  Therefore, $\kI \rightarrow \kI$ under both (a) and
(b), $\kII \rightarrow -\kII$ under both, and $\kIII \rightarrow \kIII$ under
(a) and $-\kIII$ under (b).

Now let us consider a case where one of the SIFs, before transformation,
is greater, in absolute value, for  $x_t > 0$ than for  $x_t < 0$,
so that $\ddx s {|K|} > 0$.  Then, under transformation
(a), the material at $x_t > 0$ with the greater $|K|$ moves to $x_t < 0$,
so that $\ddx s {|K|} < 0$; transformation (b) leaves $x_t$ unchanged,
so that $\ddx s {|K|} > 0$.  This can be combined with the transformation
properties of the SIFs themselves to give the transformation properties
of the gradients of the SIFs; the results are in table 1.

\tabinsert{
\begintable
Quantity \| (a)~\deg{180} $\nhat$ rotation \vb (b)~$n-t$ reflection \crthick
$\kI$  \| $+$     \vb $+$    \cr
$\kII$  \| $-$     \vb $-$    \cr
$\kIII$  \| $+$     \vb $-$    \crthick
$ \ddx s \kI$  \| $-$     \vb $+$    \cr
$ \ddx s \kII$  \| $+$     \vb $-$    \cr
$  \ddx s \kIII$  \| $-$     \vb $-$    \crthick
$ \dadotb \nhat \that   $  \| $+$     \vb $+$    \cr
$ \dadotb \nhat \bhat   $  \| $+$     \vb $-$    \cr
$ \dadotb \that \bhat   $  \| $-$     \vb $-$
\endtable
}{t1}

We also need to consider the transformation properties of gradients
of the unit vectors---quantities of the form $\dadotb {\hat a} {\hat b}$,
where $a$ and $b$ belong to $\{n,b,t\}$.  Noting that
$$\ddx s {(\hat a \cdot \hat b)} = 0 = \dadotb {\hat a} {\hat b} +
    \dadotb {\hat b} {\hat a} \putn{B1}$$
for all $a$ and $b$, since $\nhat$, $\bhat$, and $\that$
are mutually orthogonal unit vectors, we can see that there are only three
independent quantities to consider, which we take to be
$ \dadotb \nhat \that $, $ \dadotb \nhat \bhat$, and $ \dadotb \that \bhat$.
 From figure~5, we can see that the transformation properties of these three
quantities are as shown in table~1.

\pageinsert{
\hbox to 6.5 truein{
(a)
\psfig{figure=dndott.ps,width=3.0truein}
(b)
\psfig{figure=dndotb.ps,width=3.0truein}
 }
\hbox to 3.25truein{
(c)
\psfig{figure=dtdotb.ps,width=3.0truein}
}
\figcap{fig5}
}
\endinsert

\topicname{Appendix C:  Approximation of $\kII$ in two dimensions}

In this appendix, we derive equation \equn{eq9},
\equat9
which gives $\kII$ after the crack has grown a distance $x$ from the
end of a long crack of length $a$, in terms of the deviation
$\Delta\theta(x)$ of  $\theta$ (the angle that $\nhat$
makes with the $x$-axis)
from the angle that makes $\kII=0$.
First,  we can use the results of Cotterell and Rice\cite{path4}
to find that as a function of the $x$-coordinate of $\xvec$,
measured from the end of the original long crack,
$$\kII(x) = \kII(0) + \half \theta(x) \kI(0) - \left({2\over\pi}\right)^\half
     T \int_0^x {\theta(x') \over (x -x')^\half} dx' , \putn{eqc2}$$
where $T$ is the non-singular tensile stress at the end of the crack.
So, when the principle of local symmetry is satisfied, $\theta(x)$ has
the value which makes $\kII(x) = 0$;  if $\theta$ differs from this
value by a small amount $\Delta \theta(x)$, and if we take
$T = b {\kI(0) \over \sqrt a}$, with the appropriate geometrical factor $b$,
then we find that
$$\kII(x) = \kI(0) {\Delta\theta(x) \over 2} \left[ 1 - 4 b
    \left({2 x\over\pi a}\right)^\half
    \int_0^x {\Delta\theta(x') \over \Delta\theta(x)}
    \left({x \over x-x'}\right)^\half {dx'\over 2 x}\right]. \putn{eqc3}$$
Now, if ${\Delta\theta(x') \over \Delta\theta(x)}$ were constant,
then the integral on the right hand side would be equal to 1;
small variations of $\Delta\theta(x)$ from a constant function
leave the integral approximately 1.
Noting that the integral is multiplied by $\left({x\over a}\right)^\half$,
which is small by
assumption,  we can therefore approximate $\kII$ as
\equat9
which is equation \equn{eq9}.
We also note here that under the approximations of this appendix,
and with the results of Cotterell and Rice,\pcite{path4}
$\kI$  is approximately constant.

% end text:
\skipline
\centerline{\bf References}
\listrefs
%\skipline
%\skipline
%\centerline{\bf Figure Captions}
%\listfigs
%\skipline
%\skipline
\vfill\eject
\bye